\documentclass[twocolumn,showpacs,preprintnumbers,amsmath,amssymb,prl,superscriptaddress]{revtex4-1}

\usepackage{multirow}
\usepackage{amsfonts}
\usepackage{amsopn}
\usepackage[english]{babel}
\usepackage[T1]{fontenc}
\usepackage{times}
\usepackage{mathrsfs}
\usepackage{graphicx}
\usepackage{dcolumn}
\usepackage{bm}
\usepackage[colorlinks,bookmarks=true,citecolor=blue,linkcolor=red,urlcolor=blue]{hyperref}
\usepackage[tight, nooneline]{subfigure}
\subfiguretopcaptrue

	\newcommand{\ket}[1]{\left| #1 \right>}	
	\newcommand{\ketbra}[2]{\left| #1 \vphantom{#2} \right>\!\left< #2 \vphantom{#1} \right|}				
	
\begin{document}
\title{Model Fractional Chern Insulators}

\author{J\"org Behrmann}
\affiliation{Dahlem Center for Complex Quantum Systems and Institut f\"ur Theoretische Physik, Freie Universit\"at Berlin, Arnimallee 14, 14195 Berlin, Germany}
\author{Zhao Liu}
\email{zliu@zedat.fu-berlin.de}
\affiliation{Dahlem Center for Complex Quantum Systems and Institut f\"ur Theoretische Physik, Freie Universit\"at Berlin, Arnimallee 14, 14195 Berlin, Germany}
\affiliation{Department of Electrical Engineering, Princeton University, Princeton, New Jersey 08544, USA}
\author{Emil J. Bergholtz}
\email{ejb@physik.fu-berlin.de}
\affiliation{Dahlem Center for Complex Quantum Systems and Institut f\"ur Theoretische Physik, Freie Universit\"at Berlin, Arnimallee 14, 14195 Berlin, Germany}
\date{\today}

\begin{abstract}

We devise local lattice models whose ground states are \emph{model fractional
Chern insulators}---Abelian and non-Abelian topologically ordered states
characterized by exact ground state degeneracies at any finite size and infinite
entanglement gaps. Most saliently, we construct exact parent Hamiltonians for
two distinct families of bosonic lattice generalizations of the
\(\mathcal{Z}_k\) parafermion quantum Hall states: (i) color-entangled
fractional Chern insulators at band filling fractions \(\nu=k/(\mathcal{C}+1)\)
and (ii) nematic states at \(\nu=k/2\), where \(\mathcal{C}\) is the Chern
number of the lowest band. In spite of a fluctuating Berry
curvature, our construction is partially frustration free: the ground states
reside entirely within the lowest band and exactly minimize a local
\((k+1)\)-body repulsion term by term.
In addition to providing the first known models hosting intriguing states such as
higher Chern number generalizations of the Fibonacci anyon quantum Hall states,
the remarkable stability and finite-size properties make our models particularly
well-suited for the study of novel phenomena involving e.g.\ twist defects and
proximity induced superconductivity, as well as being a guide for designing
experiments.
\end{abstract}

\maketitle
\paragraph{Introduction.}
The prospect of lattice-scale fractional quantum Hall (FQH) phenomena at high
temperatures, without the need for a strong magnetic field, has attracted ample
recent attention to the theory of fractional Chern insulators
(FCIs)~\cite{otherreview,Emilreview,titusreview}.  While experimental
realizations of FCIs are becoming increasingly realistic in the light of the
recent realizations of integer Chern insulators with unit Chern number in solid
state materials~\cite{Chernexp} and cold atom systems~\cite{ChernCold}, the
theoretical frontier has turned towards strongly correlated states in bands with
higher Chern
numbers~\cite{disloc,c2,max,ChernN,dassarma,ChernTwo,ChernN2,gunnar,Grushin,coopermoessner,wu,zhao2,sunnew,simon,masa,EmilWeyl,
haldanestatforcher,Claassen,gunnarhigherc,dipolarexchange,bosonIQH,bosonIQH2}. This
is due to the fact that, although notable differences compared to the continuum
setting have been established~\cite{andreas,ifwlong,Emilreview}, all FCIs
discovered in Chern number \(\mathcal C=1\) bands have direct continuum FQH
analogs to which the adiabatic continuity has been explicitly established in several important
cases~\cite{kapit,qi,gaugefixing,AdiabaticContinuity1,AdiabaticContinuity2,AdiabaticContinuity3}.
Of special value is the early work by Kapit and Mueller who provided a natural
lattice discretization of the continuum lowest Landau level and showed that an
two-body on-site interaction leads to a perfect lattice version of the bosonic
\(\nu=1/2\) Laughlin state~\cite{kapit}. Related work has established the existence lattice parent Hamiltonians of non-Abelian states \cite{ronnyparent,annelattice,zhaononabelian}.

There is however a glaring lack of similar models describing \(\mathcal{C} > 1\)
systems, despite intriguing progress with long-ranged lattice
models~\cite{latticeC2long} and approximative mappings to continuum models with
unusual boundary conditions~\cite{wu}. Given the importance of solvable models
in the theory of topological and strongly correlated states of matter---ranging
from the AKLT model for the Haldane spin chain~\cite{AKLT} and the Kitaev chain
describing a one-dimensional \(p\)-wave superconductor~\cite{kitaev} to the
model wave functions for the continuum FQH effect~\cite{Laughlin83,mr,rr} and
their concomitant pseudopotential parent Hamiltonians~\cite{haldane83}---finding
such has remained an outstanding challenge in the theory of FCIs. This is
particularly pressing given the accumulating numerical evidence that \(\mathcal
C>1\) systems feature an even richer phenomenology than continuum Landau levels.

In the present work, we bridge this divide and provide exact lattice parent Hamiltonians
for a large class of Abelian as well as non-Abelian \emph{model} FCIs in bands
carrying any Chern number \(\mathcal C\).
We explicitly verify that the ground state multiplets are exactly degenerate at
any finite size, that the gap to excited states remains finite in the thermodynamic limit, and that there is an infinite gap in
the particle entanglement spectrum~\cite{pes,rbprx}.

\begin{figure*}
\centerline{\includegraphics[width=\linewidth]{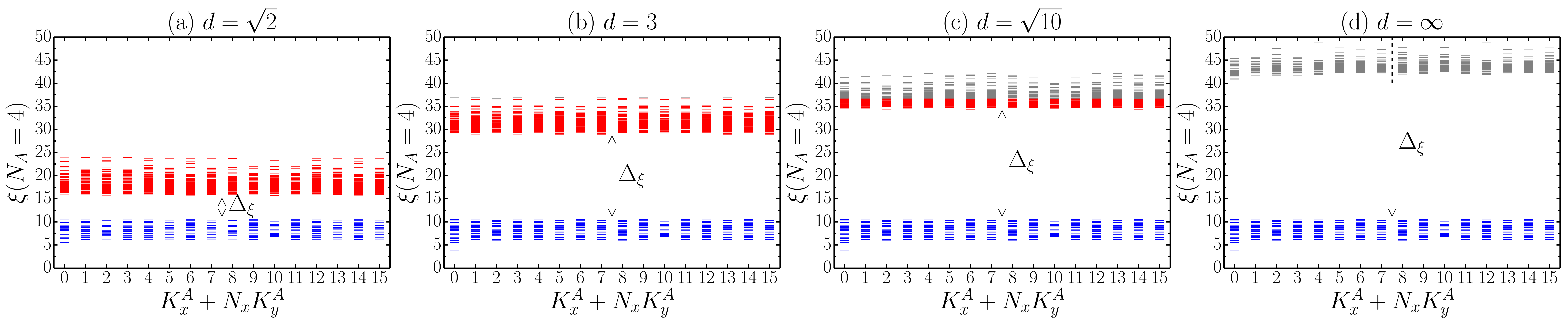}}
\caption{(Color online) Typical particle entanglement spectra (PES), here
displayed for the \(k=2\), \(\nu=2/3\) non-Abelian state in a \(\mathcal{C}=M=3\)
band with \(N=8\), \(N_x\times N_y=4\times4\), \(\phi=1/3\) and hoppings
truncated at (a) \(d=\sqrt{2}\), (b)
\(d=3\), (c) \(d=\sqrt{10}\), and (d)
\(d=\infty\) lattice constants. Generally a PES obtained by numerical
diagonalization includes three parts: the low-lying levels with quasihole
excitation information (blue), the high non-universal levels (red), and
numerical noise (gray) set by the double precision above ${\xi_c}\approx\-\ln(2^{-53})\approx36.7$.
The high, non-universal levels merge into numerical
noise for large \(d\), preventing us from further numerically tracking the
growth of the entanglement gap, which we argue to increase without bound with increasing $d$.}
\label{fig:typicalPES}
\end{figure*}

\paragraph{Flat band model.}
We begin by constructing a family of multi-orbital models possessing exactly
flat lowest bands with arbitrary Chern number \(\mathcal C\). For definitiveness
we describe our construction on a square lattice with an effective magnetic flux \(\phi=1/q\) piercing each elementary
plaquette \cite{supple}, although it can be generalized to any Bravais lattice
and rational flux.

We assign \(M\) internal orbitals to
each lattice site \(i\) with real-space coordinates
\((x_i, y_i)\), where \(M\) must be a factor of \(q\). The
site-dependent orbital index to the lattice site \(i\) is
\(s_i=x_i\bmod(\frac{q}{M})+m\frac{q}{M}\), with \(m=0,1,\dotsc,M-1\).
The single-particle physics is governed by
\begin{equation} \label{eq:hoppingham}
 H_0=\sum_{j,s_j}\sum_{k,s_k}t_{j,k}^{s_j,s_k} a^\dagger_{j,s_j}a_{k,s_k},
\end{equation}
where \(a^\dagger_{j,s_j}\) (\(a_{j,s_j}\)) creates (annihilates) a particle on
the orbital \(s_j\) at site \(j\). To achieve an exactly flat lowest Chern band
we choose the hopping amplitudes as \cite{magneto-periodic}
\begin{equation} \label{eq:hopping}
t_{j,k}^{s_j,s_k} =\delta_{s_j-x_j,s_k-x_k}^{\bmod q} (-1)^{x+y+xy} e^{-\frac{\pi}{2}(1-\phi)|z|^2} e^{-i\pi\phi(\tilde{x}_j+\tilde{x}_k)y},
\end{equation}
where \(z_j = x_j + i y_j\), \(z = z_j - z_k\), and $\tilde{x}_j=x_j+(s_j-x_j)\bmod q$. The hopping amplitudes
decay as a function of the distance between site \(k\) and site \(j\) like a
Gaussian. The hopping phase factor depends on both $x_i$ and $s_i$ according to the definition of $\tilde{x}_i$. The unit cell of our model contains \(q/M\) sites in the
\(x\) direction. The \(q\) orbitals in a unit cell lead to \(q\) bands,
and the lowest thereof is exactly flat and carries Chern number
\(\mathcal{C}=M\). For \(M=1\), our construction
Eqs.~\eqref{eq:hoppingham} and~\eqref{eq:hopping} reduces to the Kapit-Mueller
model~\cite{kapit} in Landau gauge. Similar multi-orbital models have also been
studied in Refs.~\cite{sunnew,dassarma} albeit with different choices for the
hopping amplitudes.

Although it is generally impossible to have an exactly flat band with non-zero
Chern number and strictly finite hopping~\cite{nonflat}, our model is local in
the sense of being at least exponentially bounded.
Truncating the hopping at a distance of \(d=2\) lattice constants
already gives a high flatness ratio between the band gap and
bandwidth: e.g.\ for \(\phi=1/6\) and \(\mathcal C=2 \text{ or }3\) one finds
\(f\approx 85 \text{ or } 73\) respectively. The efficiently quenched kinetic
energy amplifies the importance of interaction effects and we will now proceed to show that local
interactions indeed generate model FCIs.

\paragraph{Color-entangled FCIs.}
We begin by considering \(N\) particles with the \((k+1)\)-body on-site
repulsion on a finite lattice of \(N_x\times N_y\) unit cells with periodic
boundary conditions. The interaction Hamiltonian
reads
\begin{equation} \label{eq:onsite}
 H_{\mathrm{int}}=\sum_i \sum_{\sigma_0\leq \sigma_1\leq\dotsb\leq \sigma_k\in \{s_i\}} :\!n_{i,\sigma_0}n_{i,\sigma_1}\dotsm n_{i,\sigma_k}\!\!:,
\end{equation}
where \(n_{i,\sigma}\) is the occupation operator on the orbital \(\sigma\) at
lattice site \(i\), and \(:\!\!\dotsm\!\!:\) enforces the normal ordering. For
\(\mathcal{C}=M=1\), the single-particle wave functions of the lowest band of
Eq.~\eqref{eq:hoppingham} have the structure of a discretized lowest Landau level,
and lattice analogs of the \(\mathcal Z_k\) Read-Rezayi
states are unique zero-energy ground states of Eq.~\eqref{eq:onsite} at
\(\nu=N/(N_xN_y)=k/2\) up to an exact \((k+1)\)-fold degeneracy, when the number
of particles is a multiple of \(k\), because the exact clustering properties of
these wave functions~\cite{rr} carry over directly to the present lattice setting.
This is astonishing given that many other properties such as the fluctuating Berry
curvature in reciprocal space and the excitation spectrum
already deviate from that in the continuum since the discretized
Landau level orbitals are non-orthogonal. Furthermore, if the wave functions are
written in a properly orthogonalized Wannier
basis~\cite{qi,gaugefixing,AdiabaticContinuity3}, they differ from the continuum model
states~\cite{zhaononabelian,simon}. Nevertheless, we find that these
states are characterized by an infinite gap in the \emph{particle} entanglement
spectrum (PES) which probes the quasihole excitations of the
system~\cite{pes,rbprx}. Remarkably, we find that this scenario generalizes to any \(\mathcal{C}=M>1\): at filling fractions
\(\nu=k/(\mathcal{C}+1)\), there are \(\tbinom{\mathcal{C} + k}{k}\)-fold
exactly degenerate zero-energy ground states when the number of particles is a
multiple of \(k\), and their PES has an infinite gap.

To establish this, we project the interaction Hamiltonian
Eq.~\eqref{eq:onsite} for large number of samples onto the lowest band, and compute the many-body
eigenvalues and eigenstates by exact diagonalization. Indeed, we always
observe the expected number of zero-energy modes in the energy spectrum
\cite{note,supple}, which in turn implies that the band projection leaves the
ground states unchanged. 
We also find the number of zero-energy modes is robust
against the flux insertion (twisted boundary conditions).
To demonstrate the infinite entanglement gap in the particle entanglement spectrum
(PES), we truncate the hopping range in Eq.~\eqref{eq:hoppingham} at distance
\(d\), then track the evolution of the PES with increasing~\(d\). While the
lowest band is dispersive for finite \(d\), we study the band projected version
thus ignoring the band dispersion. For a system of \(N\) particles described by
the density matrix
\(\rho=\frac{1}{\mathcal{D}}\sum_{\alpha=1}^\mathcal{D}
\ketbra{\Psi_\alpha}{\Psi_\alpha}\), where \(\ket{\Psi_\alpha}\) is the
\(\alpha\)th state in the ground state manifold with
degeneracy~\(\mathcal{D}\)~\cite{note1}, the PES levels \(\xi\) are defined as
\(\xi\equiv-\ln\lambda\), where the \(\lambda\) are the eigenvalues of the
reduced density matrix \(\rho_A\) of \(N_A\) particles obtained by tracing out
\(N_B=N-N_A\) particles from the whole system, i.e.\ \(\rho_A=\mathrm{Tr}_B
\rho\)~\cite{rbprx}. Each PES level can be labeled by the total two-dimensional
quasi-momentum \((K_x^A,K_y^A)\) of part \(A\). When the PES levels are clearly
divided into low-lying and higher excited parts, we define the entanglement gap
as \(\Delta_\xi\equiv\xi_{i+1}-\xi_i\), where \(\xi_i\) (\(\xi_{i+1}\)) is the
highest (lowest) level in the low-lying (excited) part.

\begin{figure}
\centerline{\includegraphics[width=1.0\linewidth]{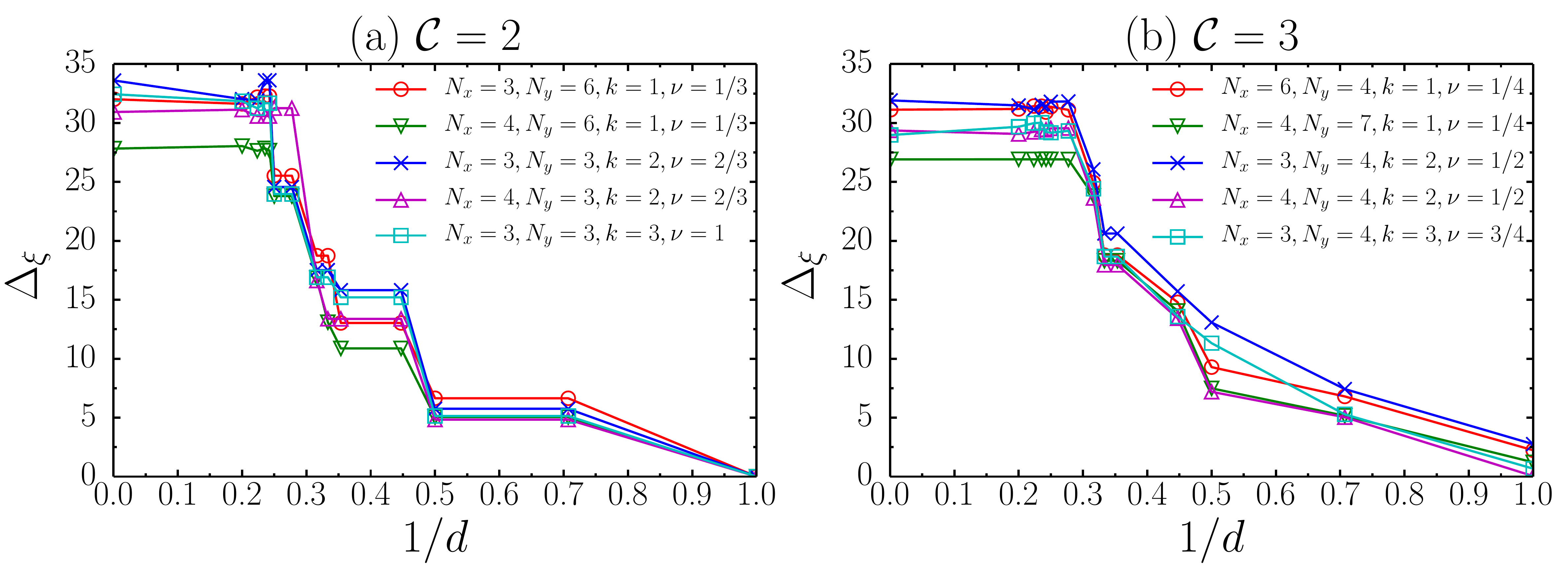}}
\caption{(Color online) The entanglement gap \(\Delta_\xi\) in the \(N_A=\lceil
N/2\rfloor\) sector versus the inverse hopping distance \(1/d\) in (a)
\(\mathcal{C}=2\) and (b) \(\mathcal{C}=3\) band. For each \(\mathcal{C}\), we
consider both Abelian states at \(\nu=1/(\mathcal{C}+1)\) and non-Abelian states
at \(\nu=2/(\mathcal{C}+1)\) and \(\nu=3/(\mathcal{C}+1)\), with lattice geometry of either
\(\gcd(N_x,\mathcal{C})=\mathcal{C}\) or \(\gcd(N_x,\mathcal{C})=1\). \(\phi\) is
chosen as \(1/\mathcal{C}\). For $d$ longer than three or four lattice
constants, the size of $\Delta_\xi$ cannot be tracked further due to the limitation
of machine precision.}
\label{entgap}
\end{figure}

Typical PES at different truncations are shown in Fig.~\ref{fig:typicalPES} for the \(\nu=1/2\) non-Abelian state in a \(\mathcal{C}=3\)
band. Including only nearest and next-nearest neighbor hopping, i.e.\ \(d=\sqrt
2\) [Fig.~\ref{fig:typicalPES}(a)], we observe a clear entanglement gap of
\(\Delta_\xi\approx5\) which is already larger than most of previously reported
results in the literature~\cite{zoology,ChernN2,ChernN}. Increasing \(d\) further elevates the non-universal part
of the PES and quickly enlarges the entanglement gap
[Figs.~\ref{fig:typicalPES}(b) and \ref{fig:typicalPES}(c)]. Our capability of tracking the
growth of the entanglement gap is only limited by the machine precision, which
determines that the PES levels can be computed reliably at most up to
\(\xi_c\approx36.7\), which corresponds to an exponentially small amplitude, of order $\mathcal O(e^{-\xi_c/2})$, in the ground state wave function. When the non-universal levels merge into the numerical
noise, it is impossible to identify the entanglement gap
[Fig.~\ref{fig:typicalPES}(d)] accurately. This happens when the machine error dominates the
high-energy part in the PES, the entanglement gap has already grown to
\(\Delta_\xi \approx 30\), which is much larger than previously reported values. We observe a similar
growth of the entanglement gap when \(1/d\rightarrow0\) in all investigated
samples, as shown in Fig.~\ref{entgap}. \(\Delta_\xi\) reaches
\(5\lesssim\Delta_\xi\lesssim7\) at \(d=\sqrt{2}\), exceeding most of the
previously reported results, and quickly increases to \(\Delta_\xi \approx 30\) at \(d\approx4\), where numerical noise starts to prevent us from further tracking the growth of \(\Delta_\xi\). However, the rapid growing of $\Delta_\xi$ and extrapolating the data to $1/d=0$ clearly suggest infinite entanglement gaps of model FCIs. A short-range truncation of
Eq.~(\ref{eq:hoppingham}) is enough to get FCIs which are essentially
indistinguishable from model FCIs with infinite entanglement gaps.

\begin{figure}
\centerline{\includegraphics[width=0.8\linewidth]{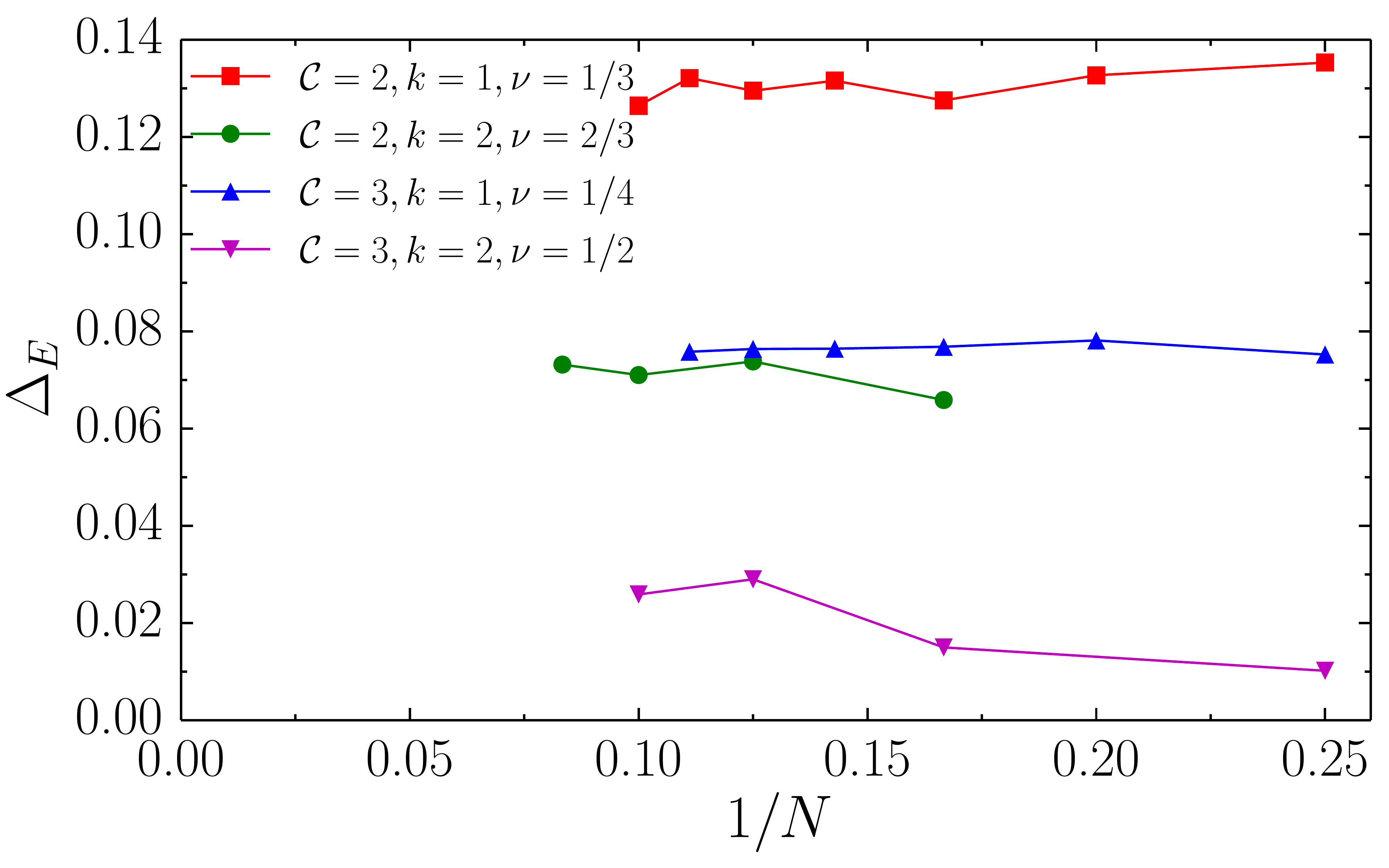}}
\caption{(Color online) The finite-size scaling of the energy gap for the
Abelian \(\nu=1/(\mathcal{C}+1)\) states and non-Abelian
\(\nu=2/(\mathcal{C}+1)\) states in \(\mathcal{C}=2\) and \(\mathcal{C}=3\)
bands. Note that the spread of the ground state manifold is exactly zero
here. \(N_x\) and \(N_y\) are appropriately chosen to make samples as isotropic
as possible. \(\phi\) is chosen as \(1/\mathcal{C}\).}
\label{engap}
\end{figure}

While the ground states and quasi-hole excitations have identically zero
interaction energy in our model, the gap \(\Delta_E\)---measured at fixed
particle number corresponding to a particle-hole excitation pair---is in
principle size-dependent. The projection of the interaction to the lowest band
will not affect the many-body gap as long as the band gap is much larger
than the interaction strength, since low-lying excitations are purely determined
by the interaction and the projection, excluding excitations caused by hopping
from lower to higher bands. In Fig.~\ref{engap}, we plot \(\Delta_E\) versus the
inverse particle number \(1/N\) for various model FCIs. In
each case we find that the gap clearly extrapolates to a finite value in the
thermodynamic limit, and, compared to other FCI models, the gap is
remarkably insensitive to the system size (cf.\ e.g.\ Ref.~\cite{andreas}).

Having established the ideal nature of FCIs in our model, we now turn to its color-entangled nature. If we interpret \(m=0,1,\dotsc,M-1\) in the orbital
indices \(s_i\) as ``layers'' or ``colors'', Eq.~\eqref{eq:hoppingham} on an infinite lattice
is equivalent to a shifted stacking of \(M\) layers of the infinite \(M=1\) model.
However, for a {\em finite} lattice of \(N_x\times N_y\) unit cells with
periodic boundary conditions, the corresponding stacking has color-entangled
boundary conditions \cite{disloc,wu,sunnew} in the \(x\)-direction in the sense that the hopping across
the boundary may occur between orbitals belonging to different layers (usual periodic boundary conditions apply in the \(y\)-direction) \cite{supple}. Crucially, each layer is not necessarily a complete
\(M=1\) model with integer number of unit cells and periodic boundary
conditions. Instead, one finds that
Eq.~\eqref{eq:hoppingham} can be mapped to \(\gcd(N_x,M)\) copies of complete
\(M=1\) model with usual periodic boundary conditions. Ref.~\cite{wu} provided a color-entangled basis built from continuum Landau levels and showed that provides a promising approach to the
\(\nu=k/(\mathcal{C}+1)\) FCIs by providing numerical evidence for a few states with small
\(\mathcal C\) and \(k\)~\cite{ChernN,ChernN2,wu}. When \(N_x\) is divisible by \(M\), the FCIs correspond to color-dependent magnetic-flux inserted versions of the Halperin~\cite{halperin} or non-Abelian
spin singlet states~\cite{NASS,NASS2}. Our construction extends this list of color-entangled FCIs and, by contrast, gives an exact construction directly in the real-space lattice.

\paragraph{Nematic states.}
The Hamiltonian Eq.~\eqref{eq:onsite} includes interactions within the same
orbital and between different orbitals.  Now we consider the zero-energy states
in the presence of only on-site \((k+1)\)-body intra-orbital repulsion, i.e.
\begin{equation} \label{eq:onsiteorbital}
H_{\mathrm{int}}=\sum_i \sum_{\sigma_0= \sigma_1=\dotsb= \sigma_k\in \{s_i\}} :\!n_{i,\sigma_0}n_{i,\sigma_1}\dotsm n_{i,\sigma_k}\!\!:.
\end{equation}
As discussed above, the single-particle problem can be mapped to
\(\gcd(N_x,M)\) copies of the \(M=1\) model. Because the interaction Hamiltonian
Eq.~\eqref{eq:onsiteorbital} does not couple different copies, the many-body
physics in this case is equivalent to distributing \(N\) on-site interacting
particles in \(\gcd(N_x,M)\) decoupled copies of \(M=1\) models, each with
\([N_x/\gcd(N_x,M)]\times N_y\) unit cells. We can count the zero-energy states
straightforwardly by this many-body mapping. For \(\nu=k/2\) with \(N\) is divisible by \(\gcd(N_x,M)\), we have
\(N/\gcd(N_x,M)\) on-site interacting particles at \(\nu=k/2\) in each copy of
the \(M=1\) model. If \(N/\gcd(N_x,M)\) is divisible by \(k\), this gives us
\(k+1\) zero-energy states obeying the same exclusion rule as the Read-Rezayi states within each
copy, and hence a total of \((k+1)^{\gcd(N_x,M)}\) zero-energy states. However, when \(N\) is not divisible by
\(\gcd(N_x,M)\), the filling fraction is larger than \(k/2\) in at least one
copy, thus there are no zero-energy states \cite{supple}. That the degeneracy depends on \(N_x\) but not on
\(N_y\) is a striking signature of the nematic nature of these
states~\cite{disloc}. Furthermore, the number of particles \(N\) does not
necessarily need to be a multiple of \(\mathcal C=M\), which further
distinguishes the nematic model FCIs from their continuum multi-layer relatives.

\paragraph{Other states.}
Following the constructions detailed above, it is straightforward to construct
parent Hamiltonians for an entire zoo of new model FCIs. For instance, for \(M=2\)
and even \(N_x\), Eq.~\eqref{eq:hoppingham} has a bilayer FQH system as the
continuum counterpart. Thus with an on-site inter-orbital three-body
repulsion in combination with a two-body intra-orbital
repulsion is expected to mimic the parent Hamiltonian for the coupled Moore-Read
state~\cite{layla,gunna} in the continuum. 
The degeneracy of this state on the torus is \(2N+3\) for even
number of particles~\cite{zliu_nr} which is consistent with our numerics~\cite{supple}. 

\paragraph{Discussion.}
In this work we have introduced \emph{model fractional Chern
insulators}---topologically ordered states with an infinite entanglement
gap---and constructed their concomitant local parent Hamiltonians directly in
the lattice.
Our construction provides natural FCI counterparts of the AKLT model for the
Haldane spin chain, the Kitaev chain and the model quantum Hall wave functions and their associated continuum
parent Hamiltonians. In analogy with these models, our construction also carries
a notion of frustration-freeness in that the ground states reside entirely within
the lowest band, and exactly minimize a strictly local \((k+1)\)-body
repulsions term by term. However, rooted in the impossibility of local Wannier functions for Chern bands \cite{hallabsence}, it appears impossible to write a gapped parent Hamiltonian, including both interactions and kinetic part, entirely as a sum of positive local terms such that each of them is minimized by the model FCIs in the two-dimensional limit, \(N_x,N_y\rightarrow\infty\)~\cite{tensorchiral,tensorchiral2,foonote}.

We have conclusively shown that our model provides an infinite
gap in the particle entanglement spectrum and that the energy gap remains finite
in the thermodynamic limit. These results are particularly remarkable
considering the strong lattice effects, e.g. reflected in a strongly non-uniform Berry curvature at small $\phi$, and underscores the incompleteness of any long-wave length description of FCIs \cite{otherreview}.

Our construction extends the list of FCIs far beyond those in the existing literature. Notably the \(k=3\)-states
(Fig.~\ref{entgap}) are the first reported \(\mathcal C>1\)
higher Chern number generalizations of the Fibonacci anyon quantum Hall states. Moreover, the nematic higher Chern number states provides a particularly promising basis for investigating lattice dislocations, which have been predicted to behave like non-Abelian wormholes in nematic Abelian parent states~\cite{disloc}. Abelian nematic states have been previously found in Refs. \cite{latticeC2long,sunnew}; our construction is giving an ideal realization in contrast to Ref. \cite{sunnew}, and is much simpler than the proposed parent Hamiltonian of Ref. \cite{latticeC2long} which includes a long tail of two-body interactions. Our non-Abelian nematic states are entirely new.

Our model also provides an ideal ground for investigating alternative platforms for Fibonacci anyons, deriving from more elementary Abelian FCIs in proximity to superconductors \cite{mong}. In particular, our $\mathcal C=2, k=1, \nu=1/3$ provides an ideal lattice version of the $(221)$-Halperin state which is a key ingredient in the construction considered in Ref. \cite{mong}.

Finally, our construction provides a guide for designing experimental
implementations of FCIs, particularly in cold atom or molecule systems
\cite{blochreview} or in arrays of qubits or nonlinear optical resonators
\cite{Hafezi,EliotLight}. The very recent observation of higher Chern numbers in photonic crystals is highly encouraging in this respect \cite{HighCExp}.

\acknowledgments
This work was supported by DFG's Emmy Noether program (BE 5233/1-1) and by the
Dahlem Research School (DRS). Z.~L. thanks Nicolas Regnault for related
work. Z.~L. was additionally supported by the Department of Energy, Office of
Basic Energy Sciences through Grant No.~DE-SC0002140.

\onecolumngrid
\section*{Supplementary Material}
In this Supplementary Material, we first discuss our flat band model with more details. We furthermore provide the ground state degeneracy (GSD), i.e.\ the number of zero modes,
obtained by numerical diagonalization of the interaction Hamiltonian for various
samples and interaction types.

\section{flat band model}
In the main text, we consider a square lattice with an effective magnetic flux $\phi=1/q$ piercing each elementary
plaquette. We assign $M$ internal orbitals to each lattice site $i$ with the real-space coordinate
$(x_i, y_i)$, where $M$ must be a factor of $q$. We then label each orbital by a site-dependent index
\begin{equation}\label{eq:si}
s_i=x_i\bmod\Big(\frac{q}{M}\Big)+m\frac{q}{M}, 
\end{equation}
with \(m=0,1,\dotsc,M-1\), such that orbitals with indices in the same residue class \(s_i
\equiv s_k \bmod (\frac{q}{M})\) share the same lattice site.
The single-particle tight-binding Hamiltonian is
\begin{equation} \label{eq:hoppingham}
 H_0=\sum_{j,s_j}\sum_{k,s_k}t_{j,k}^{s_j,s_k} a^\dagger_{j,s_j}a_{k,s_k},
\end{equation}
where \(a^\dagger_{j,s_j}\) (\(a_{j,s_j}\)) creates (annihilates) a particle on
the orbital \(s_j\) at site \(j\). To achieve an exactly flat lowest Chern band with Chern number $\mathcal{C}=M$,
we choose the hopping amplitudes as 
\begin{equation} \label{eq:hopping}
t_{j,k}^{s_j,s_k} =\delta_{s_j-x_j,s_k-x_k}^{\bmod q} (-1)^{x+y+xy} e^{-\frac{\pi}{2}(1-\phi)|z|^2} e^{-i\pi\phi(\tilde{x}_j+\tilde{x}_k)y},
\end{equation}
where \(z_j = x_j + i y_j\), \(z = z_j - z_k\), and $\tilde{x}_j=x_j+(s_j-x_j)\bmod q$. 

\begin{figure}
\centerline{\includegraphics[width=0.6\linewidth]{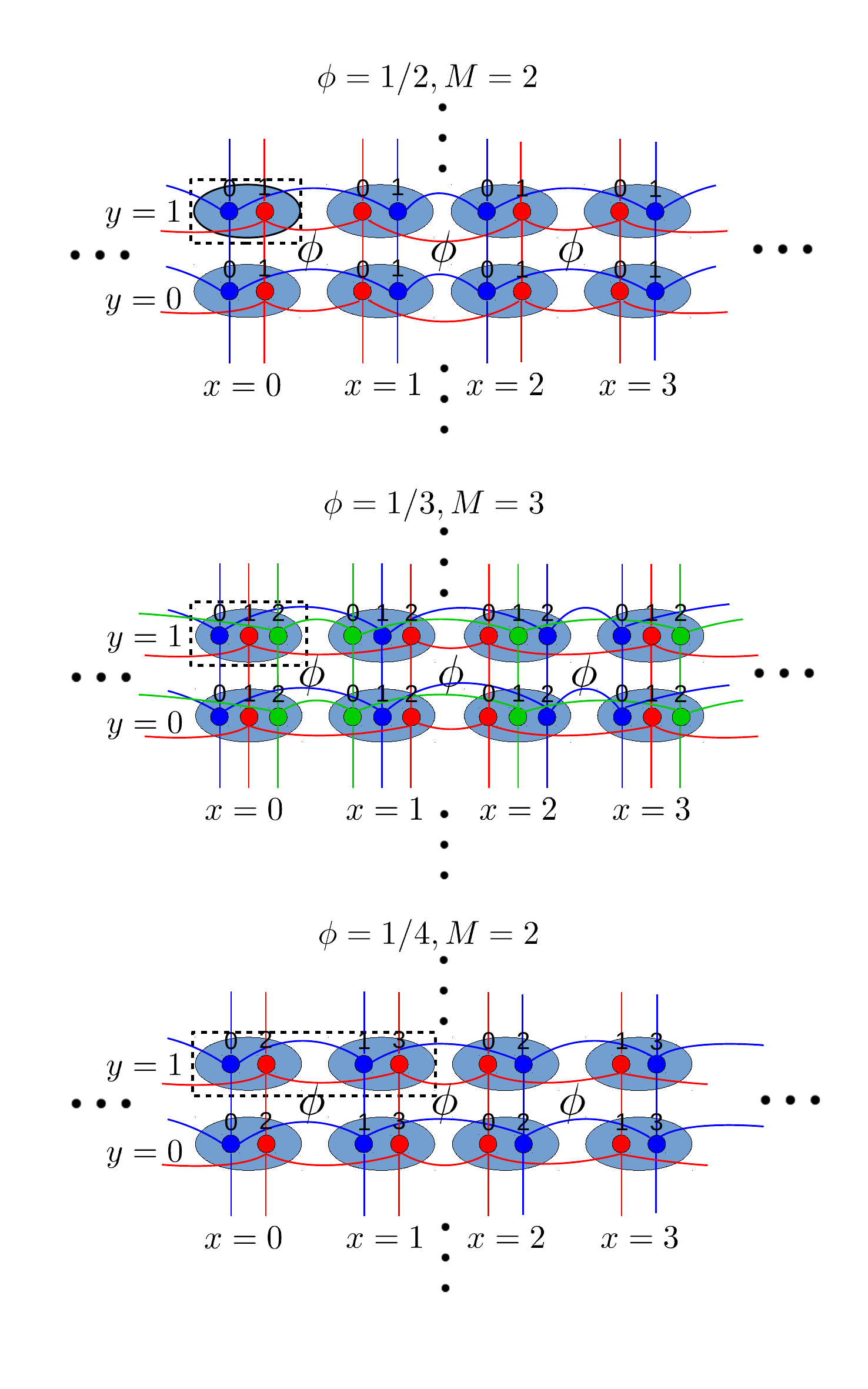}}
\caption{Typical examples of our flat band model for different $\phi=1/q$ and $M$. Each ellipse represents a lattice site with the real-space coordinate $(x,y)$. In each lattice site $i$, there are circles representing orbitals, whose indices $s_i$ are given by the numbers.
The orbitals connected by allowed hopping (only nearest-neighbor hopping is shown for simplicity) in Eq.~(\ref{eq:hopping}) have the same color and can be thought as an effective layer. The unit cell containing $q/M$ sites is indicated by the dashed rectangular. In the infinite lattice case, our model is equivalent to the stacking of $M$ layers of the $M=1$ model, with layer-dependent Aharonov-Bohm hopping phases.
}
\label{fig:lattice}
\end{figure}

In Fig.~\ref{fig:lattice}, we show typical examples of our model for $\phi=1/2,M=2$; $\phi=1/3,M=3$; and $\phi=1/4,M=2$. Once $\phi$ and $M$ are fixed, the orbital index $s_i$ in each lattice site can be easily computed from Eq.~(\ref{eq:si}). For example, for $\phi=1/4,M=2$, Eq.~(\ref{eq:si}) gives $s_i=(x_i\bmod2)+2m,m=0,1$. So we have $s_i=0,2$ for even $x_i$ and $s_i=1,3$ for odd $x_i$ (lower panel in Fig.~\ref{fig:lattice}). The unit cell, which contains $q/M$ sites, can be determined by the period of orbital indices. Hopping only occurs between orbitals satisfying $(s_j-x_j)\bmod q=(s_k-x_k)\bmod q$ due to the $\delta_{s_j-x_j,s_k-x_k}^{\bmod q}$ factor in Eq.~(\ref{eq:hopping}). If we imagine all orbitals that can be connected by hopping as an effective ``layer'', it is obvious to see that our model on an infinite lattice is equivalent to the stacking of $M$ ``layers'' of the $M=1$ model (Fig.~\ref{fig:lattice}). However, the phase of the hopping between two specific lattice sites is layer-dependent. This can be seen from the Aharonov-Bohm factor $e^{-i\pi\phi(\tilde{x}_j+\tilde{x}_k)y}$ in Eq.~(\ref{eq:hopping}), where $\tilde{x}_j$ is shifted from the site coordinate $x_j$ by an orbital (layer)-dependent term $(s_j-x_j)\bmod q$. For example, if we consider the hopping from site $(x_k,y_k)=(0,0)$ to $(x_j,y_j)=(0,1)$ in the upper panel of Fig.~\ref{fig:lattice}, a phase $e^{i 0}$ ($e^{i\pi}$) is picked up because $\tilde{x}_j=\tilde{x}_k=0$ ($\tilde{x}_j=\tilde{x}_k=1$) in the blue (red) layer. The factor $\delta_{s_j-x_j,s_k-x_k}^{\bmod q}$ in Eq.~(\ref{eq:hopping}) guarantees that the shift term $(s_j-x_j)\bmod q$ is constant in each layer. When $M=1$, we have $s_i=x_i\bmod q$ which leads to $\delta_{s_j-x_j,s_k-x_k}^{\bmod q}=1$ and $\tilde{x}_i=x_i$. Thus our model returns to the Kapit-Mueller model [30] in Landau gauge. If we only keep the 
nearest neighbor hopping, the phase shift differentiates our model from the usual multi-orbital Hofstadter model, where the hopping phase only depends on the lattice site coordinate thus no phase shift exists. Instead, our model can be thought as a multi-orbital 
Hofstadter model with the orbital-dependent hopping phase and color-entangled boundary condition.

\begin{figure}
\centerline{\includegraphics[width=0.8\linewidth]{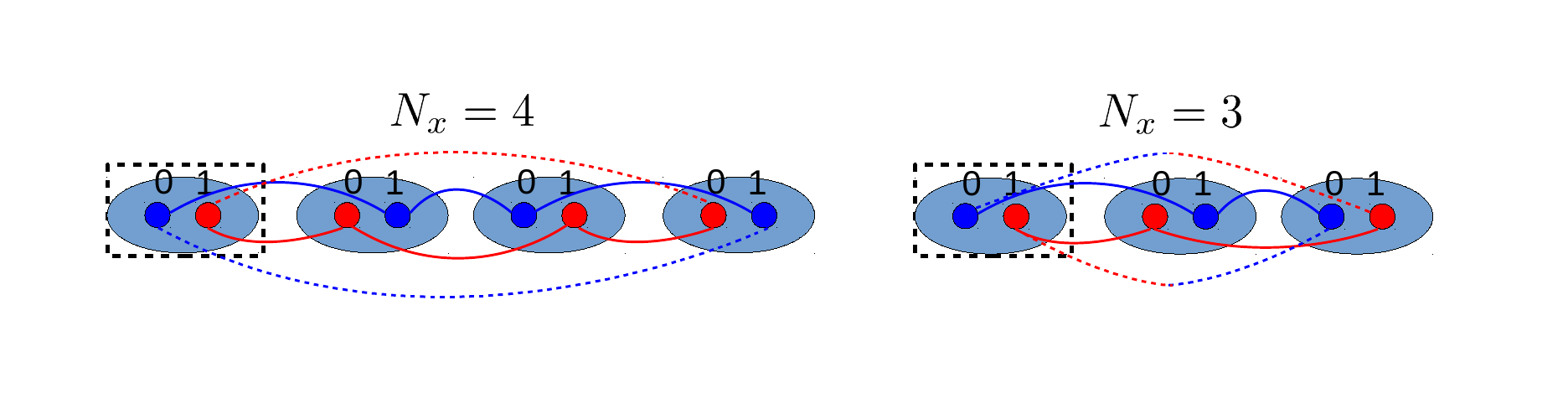}}
\caption{Our model on periodic finite lattices with $\phi=1/2$ and $M=2$. The unit cell is indicated by the dashed rectangular. The hopping across the boundary is highlighted by the dashed lines. (a) When $N_x=4$, our model is equivalent to two (blue and red) complete $M=1$ layers with periodic boundary conditions. (b) When $N_x=3$, hopping across the boundary may occur between orbitals (layers) with different colors. Thus we have color-entangled boundary conditions in the $x$-direction for the blue and red layers, and each layer is no longer a complete \(M=1\) model with integer number of unit cells and periodic boundary conditions.
}
\label{fig:latticeNx}
\end{figure}

Although it is straightforward on an {\em infinite} lattice to map our model to the shifted stacking of $M$ layers of the $M=1$ model, a {\em finite} lattice of $N_x\times N_y$ unit cells with simple periodic boundary conditions can lead to complicated boundary conditions in the layer stacking picture. Let us consider an example with $\phi=1/2,M=2$ (Fig.~\ref{fig:latticeNx}). If $N_x$ is even, by tracking the nearest-neighbor hopping in the $x$ direction, we can find that all hopping still occurs between orbitals with the same color (i.e., the same effective layer). In that case, our model is again equivalent to the shifted stacking of two complete $M=1$ layers, each of which has integer number of unit cells and periodic boundary conditions (left panel of Fig.~\ref{fig:latticeNx}). However, if $N_x$ is odd, Eq.~(\ref{eq:hopping}) leads to \(x\)-direction hopping between orbitals with different colors across the boundary (right panel of Fig.~\ref{fig:latticeNx}), which implements color-entangled boundary conditions [7,14,18] in the $x$-direction for the two effective layers. Crucially, each layer is no longer a complete \(M=1\) model with integer number of unit cells and periodic boundary conditions. Instead, by unfolding the two layers, our model is now equivalent to a single copy of $M=1$ model with usual periodic boundary conditions. In general, one can find that our model on a periodic $N_x\times N_y$ lattice can be mapped to \(\gcd(N_x,M)\) copies of complete \(M=1\) model with usual periodic boundary conditions, each copy has \([N_x/\gcd(N_x,M)]\times N_y\) unit cells and copy-dependent Aharonov-Bohm hopping phases.

\section{ground state degeneracy}
Here we summarize the ground state degeneracy, i.e.\ the number of zero modes,
obtained by numerical diagonalization of the interaction Hamiltonian for various
samples and interaction types (Tab.~\ref{t1}).
\begin{table}
\caption{The ground state degeneracy (GSD), i.e.\ the number of zero modes,
obtained by numerical diagonalization of the interaction Hamiltonian for various
samples and interaction types. As discussed in the main paper, here we consider
color-entangled FCIs, nematic FCIs, and other states (separated by the
horizontal lines). The notations of symbols are the same as in the main
paper. ``N/A'' means that we do not include the specific interaction type.}
\label{t1}

\begin{ruledtabular}
\begin{tabular}{cccccccccc}
\(N\)  & \(N_x\times N_y\) & \(M\) ($\mathcal{C}=M$) & \(\gcd(N_x,M)\) & \(\phi\) & \(\nu\) & intra-orbital interaction & inter-orbital interaction & GSD    \\
\hline
\(6\)  & \(6\times3\)      & \(2\)                   & \(2\)           & \(1/2\)  & \(1/3\) & \(k=1\)                   & \(k=1\)                   & \(3\)  \\
\(7\)  & \(3\times7\)      & \(2\)                   & \(1\)           & \(1/2\)  & \(1/3\) & \(k=1\)                   & \(k=1\)                   & \(3\)  \\
\(6\)  & \(3\times3\)      & \(2\)                   & \(1\)           & \(1/2\)  & \(2/3\) & \(k=2\)                   & \(k=2\)                   & \(6\)  \\
\(8\)  & \(4\times3\)      & \(2\)                   & \(2\)           & \(1/2\)  & \(2/3\) & \(k=2\)                   & \(k=2\)                   & \(6\)  \\
\(9\)  & \(3\times3\)      & \(2\)                   & \(1\)           & \(1/2\)  & \(1\)   & \(k=3\)                   & \(k=3\)                   & \(10\) \\
\(6\)  & \(6\times4\)      & \(3\)                   & \(3\)           & \(1/3\)  & \(1/4\) & \(k=1\)                   & \(k=1\)                   & \(4\)  \\
\(7\)  & \(4\times7\)      & \(3\)                   & \(1\)           & \(1/3\)  & \(1/4\) & \(k=1\)                   & \(k=1\)                   & \(4\)  \\
\(6\)  & \(3\times4\)      & \(3\)                   & \(3\)           & \(1/3\)  & \(1/2\) & \(k=2\)                   & \(k=2\)                   & \(10\) \\
\(8\)  & \(4\times4\)      & \(3\)                   & \(1\)           & \(1/3\)  & \(1/2\) & \(k=2\)                   & \(k=2\)                   & \(10\) \\
\(9\)  & \(3\times4\)      & \(3\)                   & \(3\)           & \(1/3\)  & \(3/4\) & \(k=3\)                   & \(k=3\)                   & \(20\) \\
\hline
\(6\)  & \(3\times4\)      & \(2\)                   & \(1\)           & \(1/2\)  & \(1/2\) & \(k=1\)                   & N/A                       & \(2\)  \\
\(8\)  & \(4\times4\)      & \(2\)                   & \(2\)           & \(1/2\)  & \(1/2\) & \(k=1\)                   & N/A                       & \(4\)  \\
\(6\)  & \(6\times2\)      & \(4\)                   & \(2\)           & \(1/4\)  & \(1/2\) & \(k=1\)                   & N/A                       & \(4\)  \\
\(7\)  & \(2\times7\)      & \(2\)                   & \(2\)           & \(1/2\)  & \(1/2\) & \(k=1\)                   & N/A                       & \(0\)  \\
\(6\)  & \(3\times2\)      & \(2\)                   & \(1\)           & \(1/2\)  & \(1\)   & \(k=2\)                   & N/A                       & \(3\)  \\
\(8\)  & \(2\times4\)      & \(2\)                   & \(2\)           & \(1/4\)  & \(1\)   & \(k=2\)                   & N/A                       & \(9\)  \\
\(10\) & \(2\times5\)      & \(2\)                   & \(2\)           & \(1/2\)  & \(1\)   & \(k=2\)                   & N/A                       & \(1\)  \\
\(9\)  & \(2\times3\)      & \(2\)                   & \(2\)           & \(1/4\)  & \(3/2\) & \(k=3\)                   & N/A                       & \(0\)  \\
\(12\) & \(2\times4\)      & \(2\)                   & \(2\)           & \(1/4\)  & \(3/2\) & \(k=3\)                   & N/A                       & \(16\) \\
\hline
\(6\)  & \(2\times6\)      & \(2\)                   & \(2\)           & \(1/2\)  & \(1/2\) & \(k=2\)                   & \(k=1\)                   & \(15\) \\
\(8\)  & \(4\times4\)      & \(2\)                   & \(2\)           & \(1/2\)  & \(1/2\) & \(k=2\)                   & \(k=1\)                   & \(19\)
\end{tabular}
\end{ruledtabular}
\end{table}

\begin{thebibliography}{99}
\bibitem{magneto-periodic}
For a finite periodic lattice, we need to extend $t_{j,k}^{s_j,s_k}$ to its magneto-periodic version.
See Ref.~\cite{kapit} for an example of $M=1$.
\bibitem{note}
For a few very small system sizes, for example \(N=2\), \(N_x \times N_y = 2
\times 2\), \(M=1\), \(k=1\), \(\phi=1/2\) and \(N=4\), \(N_x\times
N_y=2\times2\), \(M=1\), \(k=2\), \(\phi=1/2\), we observe more zero modes than
expected. These disappear by slightly increasing \(N\) or
decreasing \(\phi\).
\bibitem{note1}
Each eigenstate can be labeled by a two-dimensional
total quasi-momentum \((K_x,K_y)\). The ground state degeneracy is not always
possible to resolve for \(d=1\). In that case, we select the same number of lowest states in
each \((K_x,K_y)\) sector as the \(d=\infty\) case to construct the ground state
manifold \(\{\ket{\Psi_\alpha}\}\).
\bibitem{foonote}
Constructing quasi-local parent Hamiltonians may however be possible, but
remains a significant challenge. Moreover, if say \(N_x\) remains finite an
exponentially localised parent Hamiltonian can be constructed.
\end{thebibliography}
\end{document}